\documentclass{article}

\usepackage{arxiv}

\usepackage[utf8]{inputenc} 
\usepackage[T1]{fontenc}    
\usepackage{hyperref}       
\usepackage{url}            
\usepackage{booktabs}       
\usepackage{amsfonts}       
\usepackage{nicefrac}       
\usepackage{microtype}      
\usepackage{lipsum}
\usepackage{graphicx}
\usepackage{natbib}
\usepackage{amsmath}

\usepackage{listings}
\usepackage{xcolor}
\graphicspath{ {./Figures/} }

\title{SWEEP (Seismic Wave Equation Exploration Platform) \\A Unified Solver Framework for Differentiable Wave Physics}

\author{
 Shaowen Wang \\
  King Abdullah University of Science and Technology \\
  Thuwal, Saudi Arabia \\
  \texttt{shaowen.wang@kaust.edu.sa} \\
   \And
 Tariq Alkhalifah \\
  King Abdullah University of Science and Technology \\
  Thuwal, Saudi Arabia \\
  \texttt{tariq.alkhalifah@kaust.edu.sa}
}

\begin{document}
\maketitle

\begin{abstract}
SWEEP (Seismic Wave Equation Exploration Platform) is a unified and extensible wave equation solver library designed for wavefield modeling and inversion. It supports a wide range of wave propagation engines, including acoustic, elastic, attenuative, VTI, TTI, and their Born approximations, among others. With a built-in support for automatic differentiation, the framework enables seamless implementation of full-waveform inversion (FWI), least-squares reverse time migration (LSRTM), and other gradient-based optimization methods. It also features a plug-and-play architecture, allowing easy integration and flexible combination of custom loss functions, multi-GPU computation, neural networks, and more. This makes Sweep a powerful and customizable platform for tackling advanced seismic inverse problems.\\
\end{abstract}

\section{Introduction}
Wave-equation-based velocity model building and imaging are fundamental tasks in seismic exploration. These processes typically involve solving wave equations that govern the propagation of seismic waves through the Earth's subsurface. For inverse problems, the corresponding adjoint wave equations must also be solved. Traditionally, this requires manually deriving the adjoint equations and associated gradient expressions using the adjoint-state method \citep{virieux2009overview}. However, deriving and implementing these adjoint wave equations can be complex and demands careful attention to wave physics, including the temporal consistency between forward and backward wavefields (the dot product test, \cite{claerbout1985imaging}), boundary conditions, and source terms. This complexity often presents challenges in maintaining code quality, ensuring correctness, and achieving computational efficiency.

Automatic differentiation (AD) provides a powerful alternative to the manual derivation and implementation of adjoint wave equations. By leveraging AD, we can compute gradients of loss functions with respect to model parameters—or wavefields—as long as the involved operations are differentiable. The current challenge of developing AD-based solvers lies in balancing implementation simplicity \citep{seistorch, adfwi} with computational performance \citep{deepwave}: PyTorch-based implementations offer high development efficiency but often suffer from lower runtime performance, while CUDA-based implementations provide high computational efficiency at the cost of increased complexity and reduced development flexibility. In this context, the emergence of JAX \citep{jax2018github} presents a promising solution to this dilemma. JAX offers automatic differentiation capabilities similar to PyTorch but combines them with just-in-time (JIT) compilation to achieve performance comparable to hand-written CUDA codes. Built on the XLA compiler, JAX can optimize and compile Python functions into high-performance machine code, enabling efficient execution on GPUs and TPUs.

Building on PyTorch and JAX libraries, we developed SWEEP (Seismic Wave Equation Exploration Platform), which supports both frameworks as backends for the same wave equation solver. It is also a repository with a clean and concise codebase—particularly in its wave equation implementation—that significantly enhances usability and flexibility. Its streamlined and efficient solver design allows users to easily implement higher-level operations such as wavefield separation, common image gather (CIG) computation, and other advanced processing tasks. Furthermore, it enables seamless integration with diverse workflows, including neural-network-based velocity representations, multi-scale inversion strategies, source encoding techniques, and optimized loss function designs. By reducing the complexity of the low-level complexity operations, SWEEP empowers users to concentrate on high-level workflow designs necessary to address the underlying scientific challenges.

\section{Features}
 
In Sweep, wave equations are primarily implemented using the time-domain finite-difference (FDTD) method, with a focus on 2D simulations. A central component of the framework is the time-stepping scheme, which updates the wavefield at each time step based on its previous values and the spatial derivatives defined by the underlying physical model.

While the implementation follows standard finite-difference techniques, the time-stepping process is abstracted in a way that is structurally similar to a recurrent neural network (RNN) \citep{wavetorch, sun2020theory}. This analogy serves as a conceptual abstraction rather than a functional equivalence—the propagation is not carried out using neural networks, but the recurrent formulation offers a unified and extensible interface for incorporating various wave equations. Importantly, this abstraction is not limited to supporting different physical models (e.g., acoustic, elastic, viscoacoustic); it also provides standardized interfaces for common operations such as source injection and data recording.

This modular design enhances code reusability, extensibility, and compatibility with automatic differentiation frameworks, enabling flexible integration into workflows such as full-waveform inversion (FWI), least-squares reverse time migration (LSRTM), and learning-based approaches.

\subsection{Flexible implementation}
There are three main modules in the Sweep wave equation implementation: \textbf{Propagator}, \textbf{Source}, and \textbf{Receiver}. The \textbf{Propagator} module is responsible for the time-stepping process, handling memory allocation and wavefield updates at each time step. The \textbf{Source} module manages the injection of wavefields into the simulation, supporting flexible source configurations. The \textbf{Receiver} module records wavefields at specified locations and for specified components, enabling data collection for analysis or inversion tasks.

To implement a specific wave equation, users only need to define the wavefield update behavior for a single time step. There is no need to manage model shapes or memory allocation—these are automatically inferred from the input model dimensions and configurations. The \textbf{Propagator} module orchestrates the full time-stepping process based on the provided parameters, and, in combination with the \textbf{Source} and \textbf{Receiver} modules, automatically executes the complete simulation.

The following code snippet demonstrates how we can numerically solve a simple acoustic wave equation in \textbf{Sweep}:

\begin{lstlisting}
def step(u_now, u_pre, vp, dt, h, b, lap_u_now):    
    u_next = 2 * u_now - u_pre + vp**2 * dt**2 * lap_u_now
    return u_next, u_now
class Acoustic:

    def __init__(self, ...):
        ...

    @property
    def models(self):
        return ['vp']
    
    @property
    def wavefields(self):
        return ['h1', 'h2']

    def func(self, *args, **kwargs):
        lap_u_now = laplace(args[0], args[4], self.kernel)
        return step(*args, lap_u_now)

\end{lstlisting}

The $step$ function defines the wavefield update for a single time step, while the $Acoustic$ class encapsulates the entire wave equation implementation. The $models$ property specifies the required physical model parameters (e.g., velocity), and the $wavefields$ property defines symbolic names for the wavefields involved in the simulation. These symbolic names indicate how many wavefields are needed and serve as keys for injecting sources and recording data. The $func$ method implements the time-stepping logic, including the computation of spatial derivatives using a Laplacian operator.

\subsection{Batch modeling}
To maximize computational efficiency, a batch simulation approach is provided. This allows multiple shots to be computed simultaneously, reducing the number of kernel launches compared to processing individual shots sequentially. Assume that we need to launch the forward modeling kernel $k$ times for a single source. If we model $n$ shots sequentially, the kernel must be launched $n \times k$ times. However, with batch modeling, we only need to launch the kernel $k$ times, once for all shots, significantly reducing kernel launch overhead and improving overall performance. In addition, this approach enables simulations with multiple different velocity models simultaneously, supporting more diverse and flexible simulation scenarios.

\subsection{High performance computing}
Sweep is designed to leverage high-performance computing (HPC) resources, including GPUs and TPUs. Although the underlying wave equation implementation targets single-device execution, it can be easily extended to multiple GPUs using PyTorch’s $DistributedDataParallel$ wrapper or JAX’s $pmap$, enabling users to effectively exploit shot-based parallelism. Combined with batch modeling, this approach allows for efficient simulations across multiple devices, significantly accelerating the computation process. With minimal modifications, users can extend a single-device program to run seamlessly on multiple devices.

\begin{figure}[h!]
\centering
\includegraphics[width=0.3\textwidth]{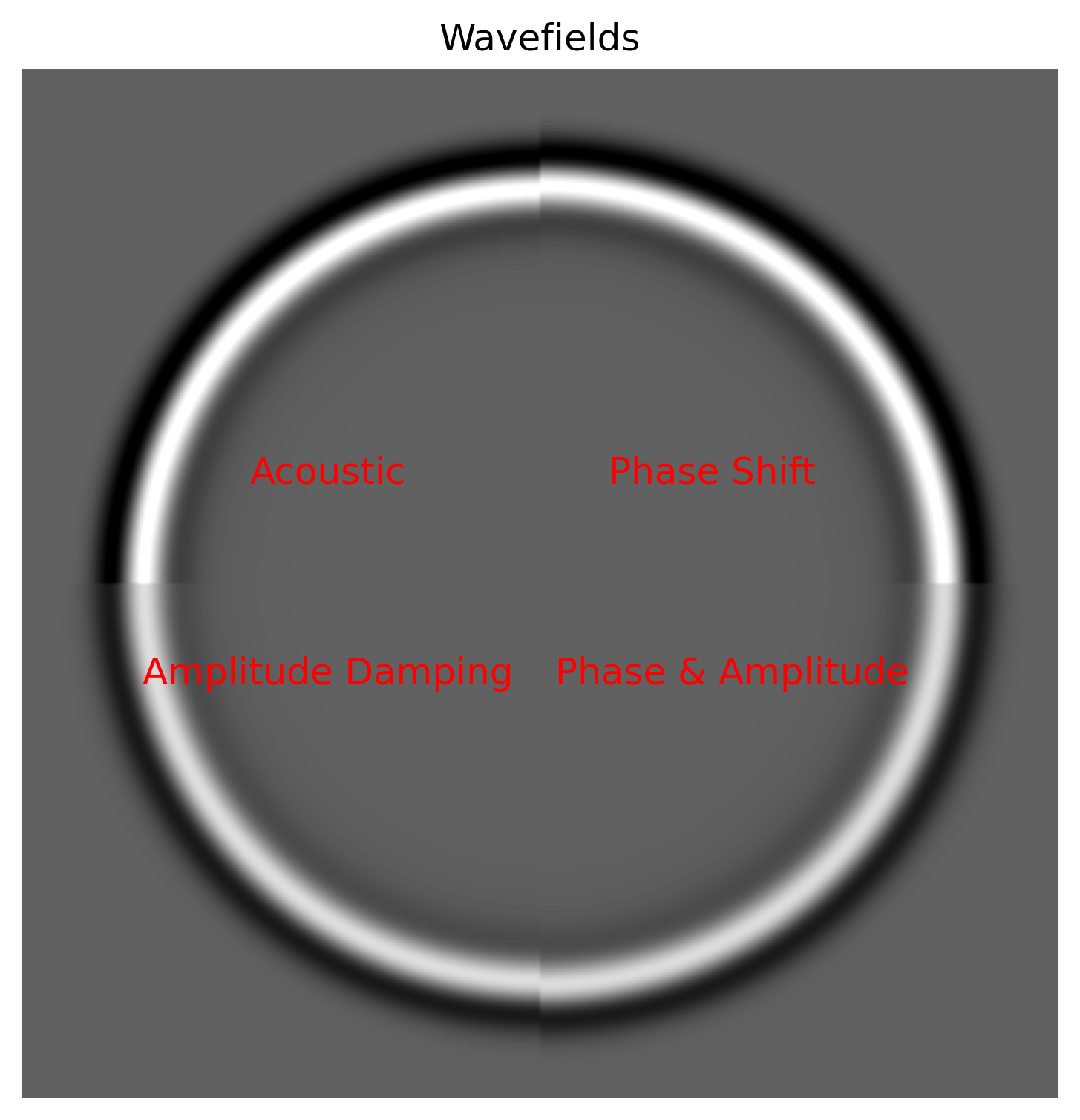}
\caption{Wavefields of the viscoacoustic wave equations.}
\label{fig:viscoacoustic_wavefields}
\end{figure}

\begin{figure}[h!]
\centering
\includegraphics[width=0.7\textwidth]{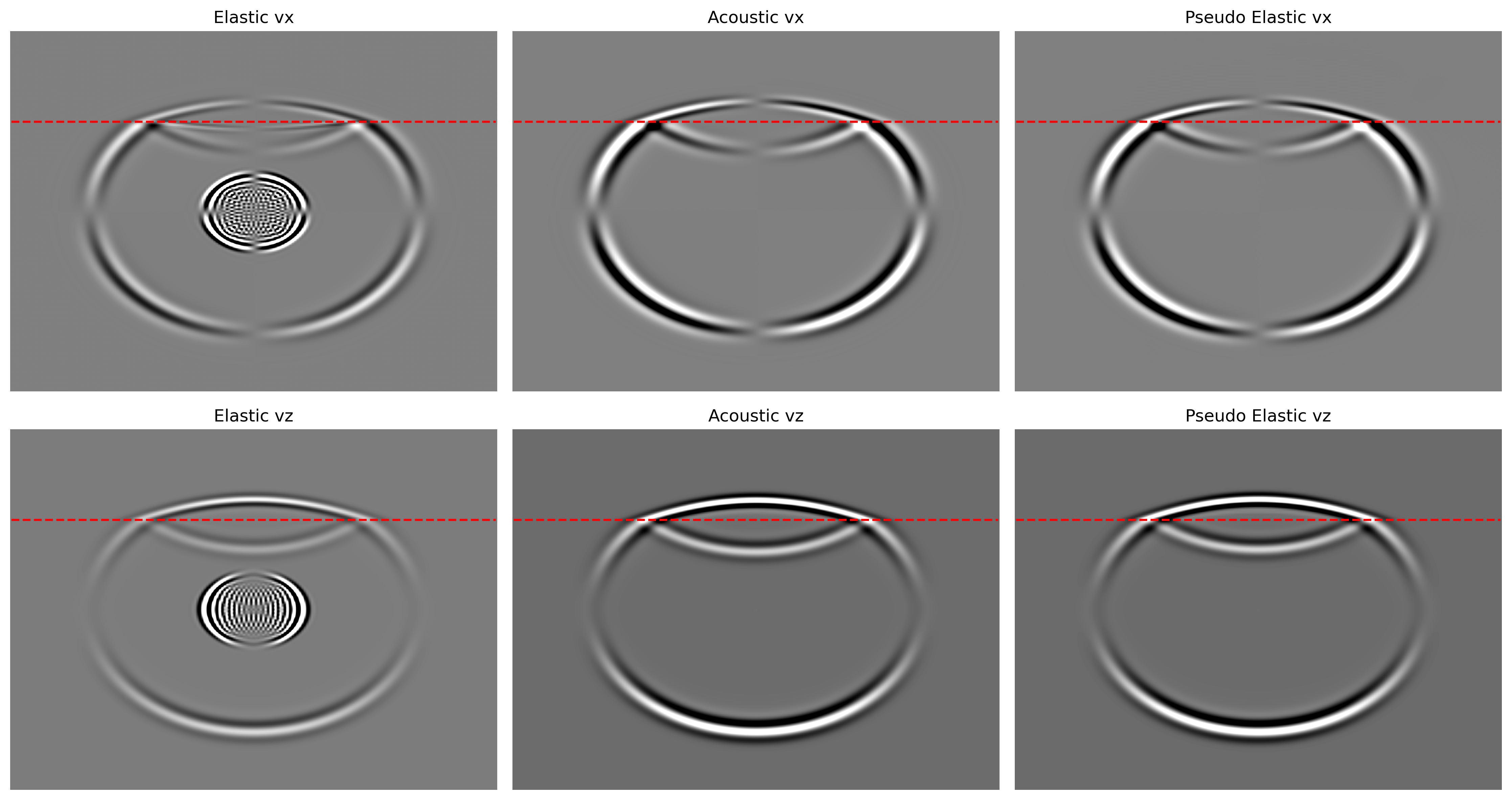}
\caption{Wavefields of the pseudoelastic wave equations.}
\label{fig:pseudoelastic_wavefields}
\end{figure}

\begin{figure}[h!]
\centering
\includegraphics[width=0.7\textwidth]{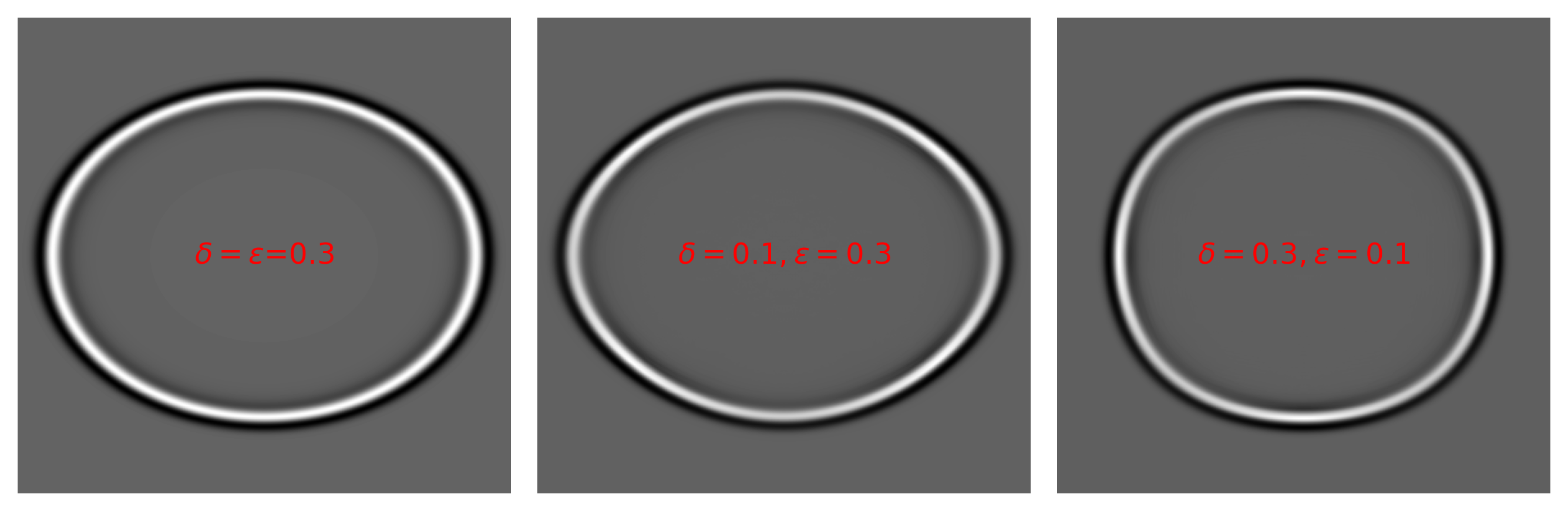}
\caption{Wavefields of the VTI acoustic wave equation with different anisotropic parameters.}
\label{fig:vti_wavefields}
\end{figure}

\begin{figure}[h!]
\centering
\includegraphics[width=0.7\textwidth]{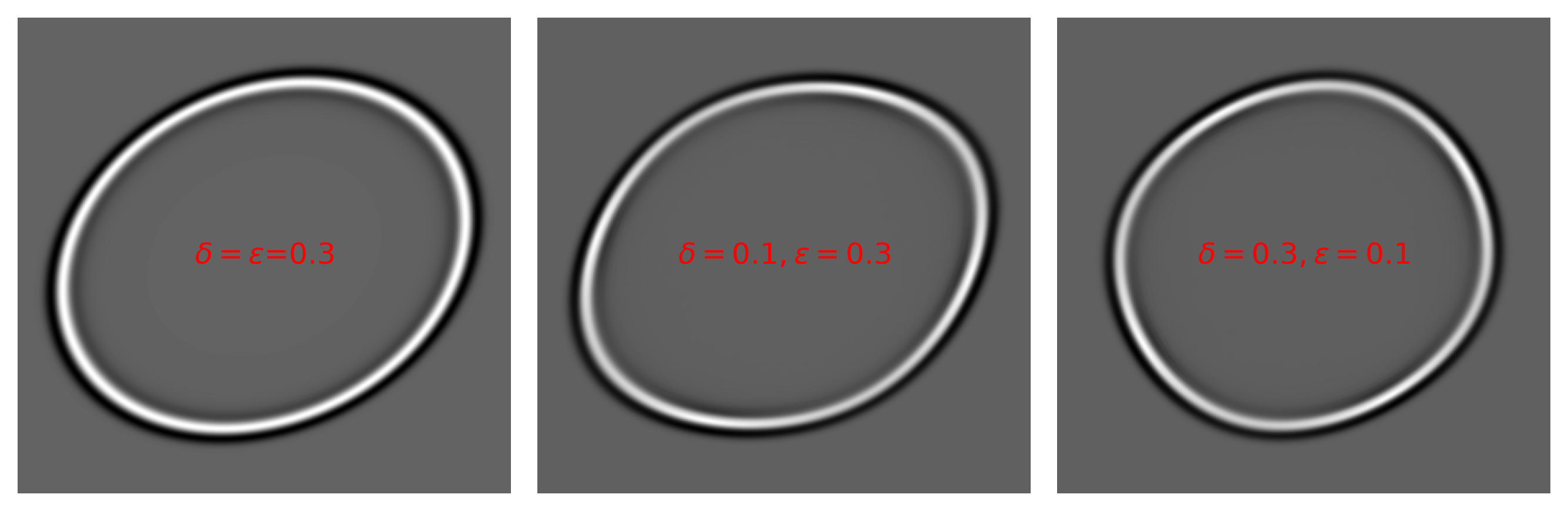}
\caption{Wavefields of the TTI acoustic wave equation with different anisotropic parameters.}
\label{fig:tti_wavefields}
\end{figure}

\section{Equation Examples}

\subsection{Acoustic Wave Equation}
\subsubsection{First order acoustic wave equation}
The first-order acoustic wave equation can be expressed as
\begin{equation}
\begin{aligned}
&\frac{\partial P}{\partial t} = -\rho v_p^2  (\frac{\partial v_x}{\partial x} + \frac{\partial v_z}{\partial z}), \\
&\frac{\partial v_x}{\partial t} = -\frac{1}{\rho} \frac{\partial P}{\partial x}, \\
&\frac{\partial v_z}{\partial t} = -\frac{1}{\rho} \frac{\partial P}{\partial z},
\end{aligned}
\end{equation}
where $\mathbf{P}$ is the pressure wavefield, $\mathbf{v} = (v_x, v_z)$ is the particle velocity, $\rho$ is the density, and $v_p$ is the P-wave velocity. This equation can be solved by using a staggered grid finite difference method \citep{acoustic1st}.

\subsubsection{Second order acoustic wave equation}
The second-order acoustic wave equation under the constant density assumption can be expressed as
\begin{equation}
\begin{aligned}
&\frac{\partial^2 P}{\partial t^2} = v_p^2 \left( \frac{\partial^2 P}{\partial x^2} + \frac{\partial^2 P}{\partial z^2} \right),\\
\end{aligned}
\end{equation}
where $\mathbf{P}$ is the pressure wavefield and $v_p$ is the P-wave velocity. Typically, this equation is solved using a uniform grid finite difference method.

\subsubsection{Example}
\begin{figure}[h!]
\centering
\includegraphics[width=0.5\textwidth]{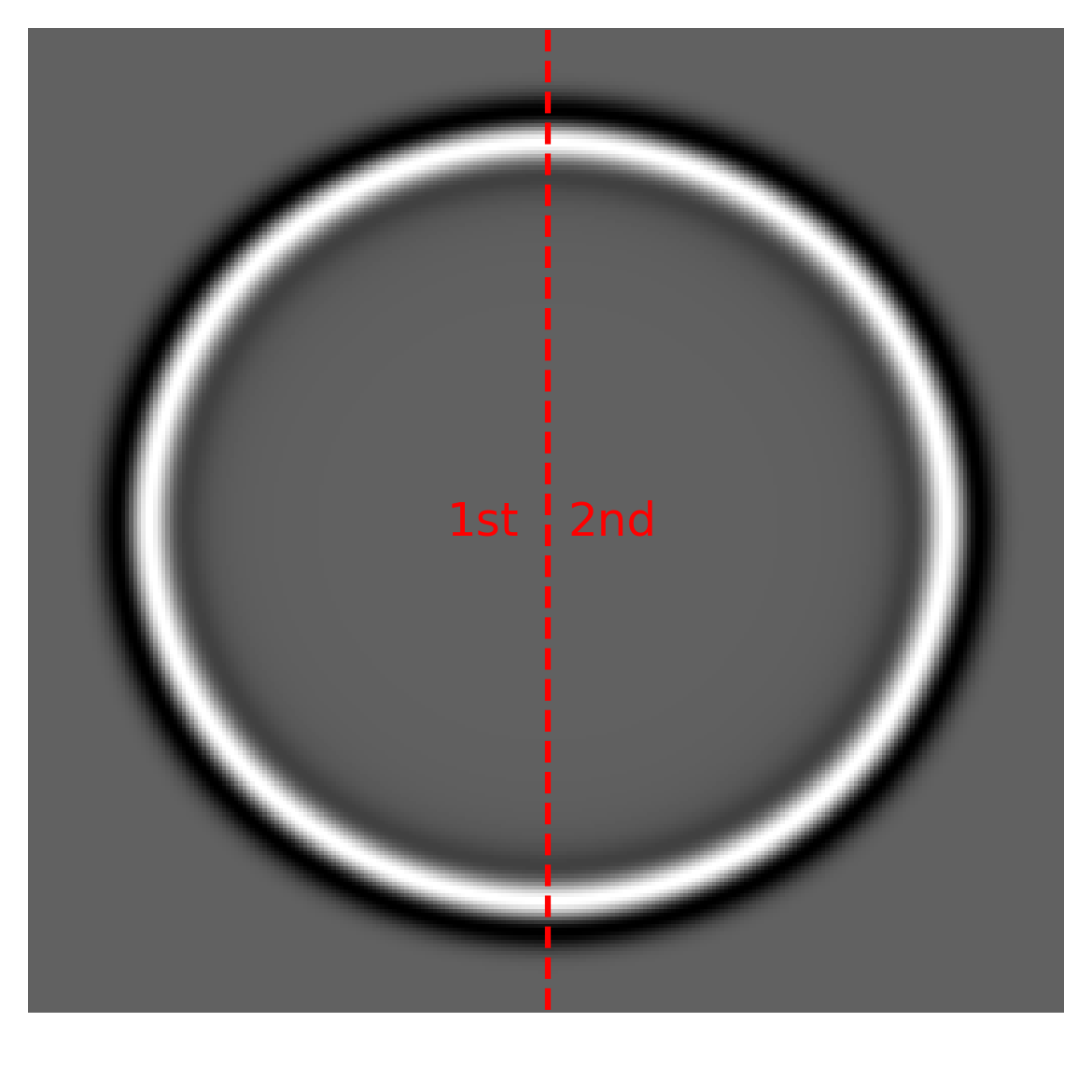}
\caption{Wavefields of the first order and second acoustic wave equation.}
\label{fig:acoustic_wavefields}
\end{figure}

The P-component of the first-order acoustic wave equation can be derived from the second-order acoustic wave equation by integrating it over time. Figure~\ref{fig:acoustic_wavefields} shows the wavefields for both the first-order and second-order acoustic wave equations. To generate the same wavefields, the source for the first-order equation is an integrated Ricker wavelet, while the source for the second-order equation is a Ricker wavelet with a peak frequency of $20 Hz$. The model is a 2D homogeneous medium with a P-wave velocity of 2000 $m/s$ and a density of 1000 $kg/m^3$. The source is placed at the center of the model, and the wavefields are displayed at different time steps.

\subsection{Acoustic with vector reflectivity}
The acoustic wave equation with vector reflectivity \citep{vf} can be expressed as
\begin{equation}
\begin{aligned}
&\frac{\partial^2 P}{\partial t^2} = v_p^2 \nabla^2 P + v_p \nabla v_p \cdot \nabla P - 2 v_p^2(\mathbf{R} \cdot \nabla P),
\end{aligned}
\end{equation}
where $\mathbf{P}$ is the pressure wavefield, $v_p$ is the P-wave velocity, and $\mathbf{R}=(r_x, r_z)$ is the vector reflectivity. We can also use a impedance reflectivity $Z$ to represent this equation with the relation $\mathbf{R}=\frac{1}{2}\frac{\nabla Z}{Z}=-\frac{1}{2}Z\nabla(\frac{1}{Z})$
\begin{equation}
\begin{aligned}
&\frac{\partial^2 P}{\partial t^2} = v_p^2 \nabla^2 P + v_p \nabla v_p \cdot \nabla P + v_p^2Z \nabla(\frac{1}{Z} \cdot \nabla P),
\end{aligned}
\end{equation}

\subsection{Anisotropy Acoustic Wave Equation}

\subsubsection{VTI}
The acoustic wave equation for 2D vertical transversed isotropic (VTI) media \citep{qPtariq} can be expressed as
\begin{equation}
\begin{aligned}
&\frac{\partial^2 P}{\partial t^2} = (1+2\eta)v^2
\frac{\partial^2 P}{\partial^2 x}+v_v^2\frac{\partial^2 P}{\partial^2 z}-2\eta v^2 v_v^2 (\frac{\partial^4 F}{\partial^2 x \partial^2 z}), \\
&P=\frac{\partial^2 F}{\partial t^2},
\end{aligned}
\end{equation}
where $v$ is the interval NMO velocity, $v_v$ is the vertical P-wave velocity, and $\eta$ is the anisotropy parameter. The differential scheme can be found in equations 24-27 of \cite{qPtariq}.

\subsubsection{Example}
\begin{figure}[htbp]
\centering
\includegraphics[width=0.8\textwidth]{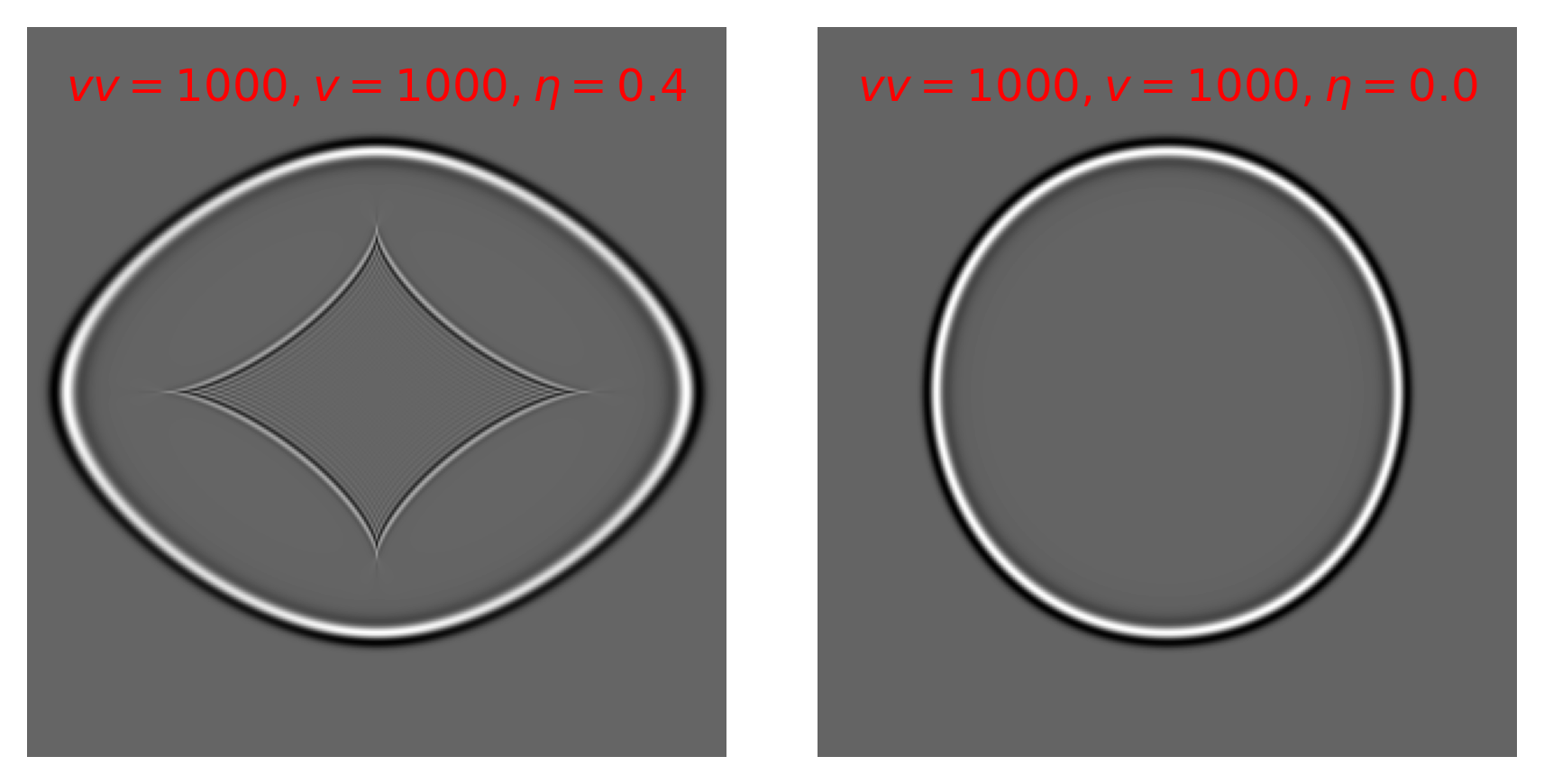}
\caption{Wavefields of the VTI acoustic wave equation (left: $\eta=0.4$, right: $\eta=0.0$).}
\label{fig:wavefields_tariq}
\end{figure}

A model with $v_v$=1000 $m/s$, $v$=1000 $m/s$, and $\eta=0.4$ and $\eta=0.01$ are used for modeling. The source is a Ricker wavelet with a peak frequency of 10 Hz. The qP wavefields generated by this equation has artifacts with a faint diamond pattern (Figure \ref{fig:wavefields_tariq} left)\citep{qPtariq} while the isotropic medium (Figure \ref{fig:wavefields_tariq} right) does not have this pattern. The wavefields are shown in Figure \ref{fig:wavefields_tariq}.

\subsubsection{Decoupled VTI}
The decoupled acoustic wave equation in VTI media \citep{qPliang} can be expressed as
\begin{equation}
\frac{\partial^2 P}{\partial t^2}=V_{\mathrm{P} 0}^2\left[(1+2 \varepsilon)+S_k\right] \frac{\partial^2 P}{\partial x^2}+V_{\mathrm{P} 0}^2\left(1+S_n\right) \frac{\partial^2 P}{\partial z^2}
\end{equation}
where $P$ is the qP-wave, $V_{\mathrm{P} 0}$ is the P-wave velocity, and $S_k$ can be calculated by:
\begin{equation}
S_n=\frac{-2(\varepsilon-\delta)(\frac{\partial P}{\partial x})^2(\frac{\partial P}{\partial z})^2} {(1+2 \varepsilon) (\frac{\partial P}{\partial x})^4+(\frac{\partial P}{\partial z})^4+2(1+\delta) (\frac{\partial P}{\partial x})^2 (\frac{\partial P}{\partial z})^2}
\end{equation}
where $\varepsilon$ and $\delta$ are the anisotropy parameters. The differential scheme can be found in equation 21 of \cite{qPliang}.

\subsubsection{Example}

\begin{figure}[h!]
\centering
\includegraphics[width=1.0\textwidth]{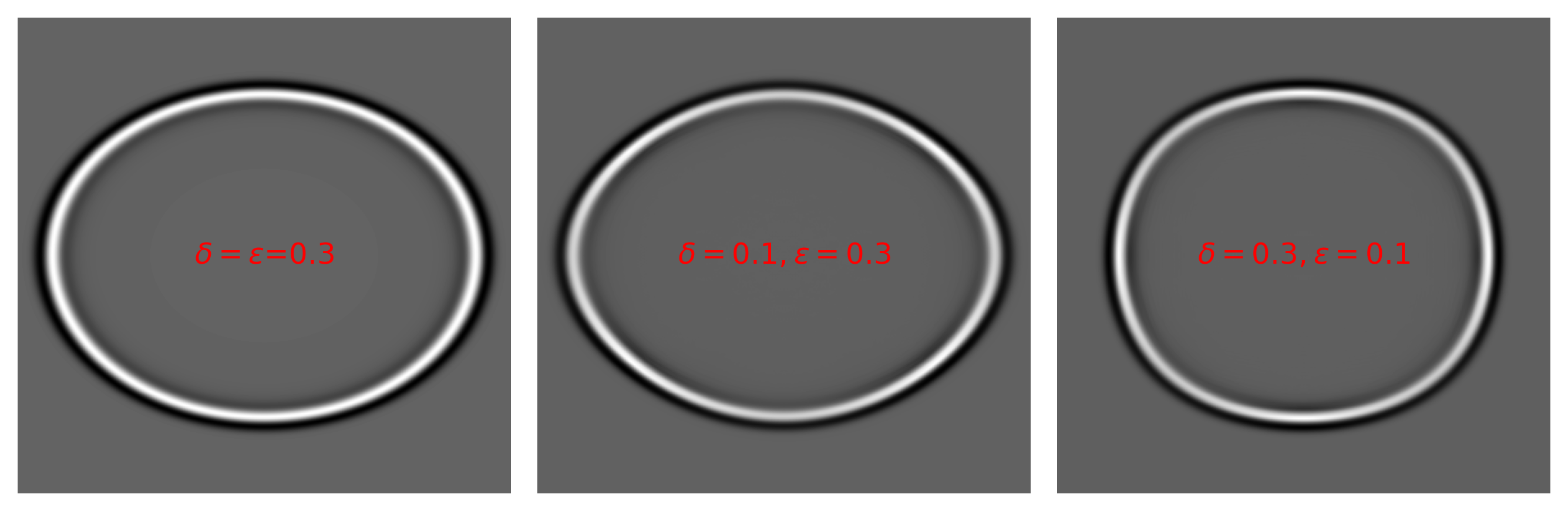}
\caption{Wavefields of the VTI acoustic wave equation with different anisotropic parameters.}
\label{fig:liang_vti}
\end{figure}

Figure \ref{fig:liang_vti} shows the wavefields generated by the decoupled VTI acoustic wave equation with different anisotropic parameters. The left shows the wavefields with $\delta=\epsilon=0.3$, the middle shows the wavefields with $\delta=0.1,\epsilon=0.3$, and the right shows the wavefields with $\delta=0.3,\epsilon=0.1$. We can see that there's no degenerate qSV artifacts even $\epsilon > \delta$.

\subsubsection{Decoupled TTI}
The decoupled acoustic wave equation in VTI media \citep{qPliang} can be expressed as

\begin{equation}
\begin{aligned}
\frac{\partial^2 P}{\partial t^2} &= V_{\text{P0}}^2 \left[ (1 + 2\varepsilon)\cos^2\theta^0 + \sin^2\theta^0 + S_k \right] \frac{\partial^2 P}{\partial x^2} \\
&+ V_{\text{P0}}^2 \left[ (1 + 2\epsilon)\sin^2\theta^0 + \cos^2\theta^0 + S_k \right] \frac{\partial^2 P}{\partial z^2} \\
&- 2\epsilon V_{\text{P0}}^2 \sin(2\theta^0) \frac{\partial^2 P}{\partial x \partial z}.
\end{aligned}
\end{equation}

where $P$ is a scalar wavefield, $V_{\text{P0}}$ is the P-wave velocity, $\theta^0$ is the angle in TTI media, and $S_k$ can be calculated by:

\begin{equation}
\begin{aligned}
S_k =\ & -2(\varepsilon - \delta)(k_x \cos\theta^0 - k_z \sin\theta^0)^2 (k_x \sin\theta^0 + k_z \cos\theta^0)^2 \cdot \frac{1}{\text{Deno}}, \\
\text{Deno} =\ & (1 + 2\varepsilon)(k_x \cos\theta^0 - k_z \sin\theta^0)^4 + (k_x \sin\theta^0 + k_z \cos\theta^0)^4 \nonumber \\
& + 2(1 + \delta)(k_x \cos\theta^0 - k_z \sin\theta^0)^2 (k_x \sin\theta^0 + k_z \cos\theta^0)^2.
\end{aligned}
\end{equation}
where $\varepsilon$ and $\delta$ are the anisotropy parameters, and $k_x$ and $k_z$ are the wavenumber components in the $x$ and $z$ directions, respectively.

\subsubsection{Example}

\begin{figure}[h!]
\centering
\includegraphics[width=1.0\textwidth]{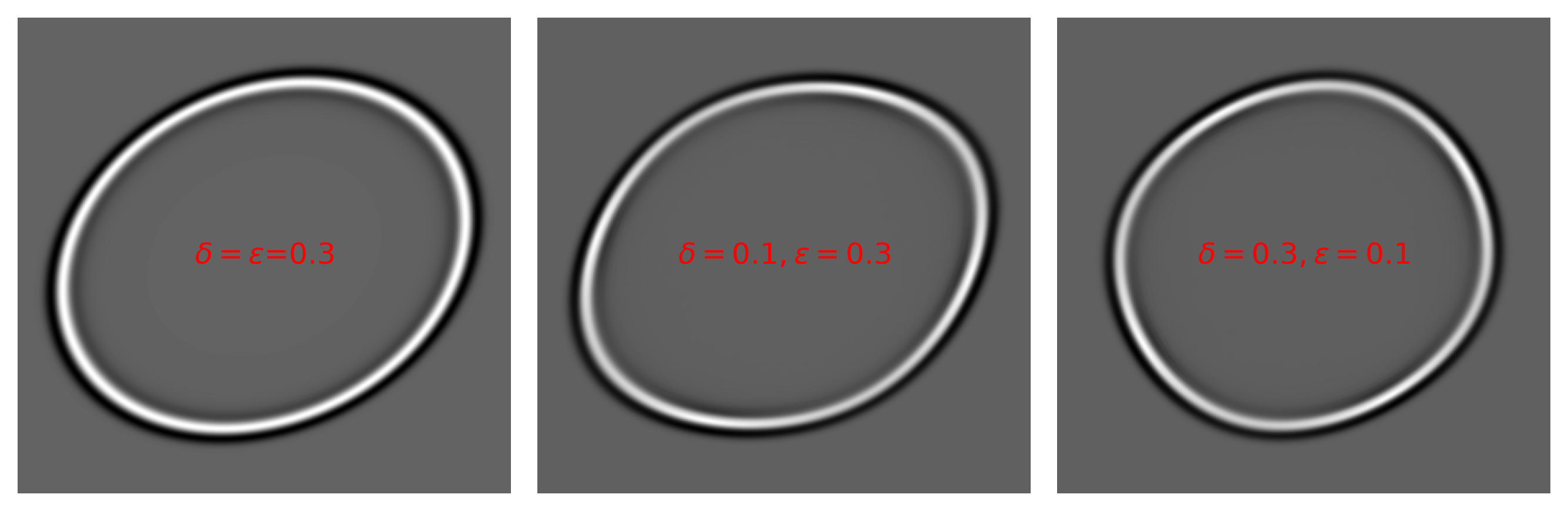}
\caption{Wavefields of the TTI acoustic wave equation with different anisotropic parameters.}
\label{fig:liang_vti}
\end{figure}

Figure \ref{fig:liang_vti} shows the wavefields generated by the decoupled TTI acoustic wave equation with different anisotropic parameters, the settings of $\epsilon$ and $\delta$ are the same as the VTI case. The angle $\theta$ is set to 30.
\subsection{Viscoacoustic Wave Equation}
The viscoacoustic wave equation based on the second order acoustic wave equation can be expressed as
\begin{equation}
\begin{aligned}
\frac{\partial^2 u}{\partial t^2} &= v_p^2 \nabla^2 u 
- \underbrace{( (1 - \sqrt{1 + Q^{-2}} ) Q^{-2} ) v_p^2 \nabla^2 u}_{\text{Phase shift correction}} \\
&\quad - \underbrace{ ( \frac{t v_p}{2} ) \mathcal{F}^{-1} [ \mathbf{k} \cdot \mathcal{F} ( \frac{\partial u}{\partial t} ) ] }_{\text{Amplitude damping}}
\end{aligned}
\end{equation}
where $u$ is the pressure wavefield, $v_p$ is the P-wave velocity, $Q$ is the quality factor, and $\mathcal{F}$ denotes the Fourier transform, $\mathcal{F}^{-1}$ denotes the inverse Fourier transform, and $\mathbf{k}$ is the wavenumber vector. Given frequency $\omega$ and $Q$, the intermediate variable $t$ can be calculated as

\begin{equation}
\begin{aligned}
    t_\sigma &= \frac{1}{\omega} \left( \sqrt{1 + \frac{1}{Q^2}} - \frac{1}{Q} \right), \\
t_\epsilon &= \frac{1}{\omega^2 t_\sigma}, \\
t &= \frac{t_\epsilon}{t_\sigma - \varepsilon} - 1, \quad \text{where } \varepsilon = 10^{-8}.
\end{aligned}
\end{equation}

\subsubsection{Example}

A model with a constant P-wave velocity of 2000 $m/s$ and a quality factor $Q$ of 50 is used to illustrate the effects of attenuation and dispersion in acoustic wave equations. The source is a Ricker wavelet with a frequency of 20 Hz, and the angular frequency $\omega$ is set to 40 $\pi$.

\begin{figure}[htbp]
\centering
\includegraphics[width=0.5\textwidth]{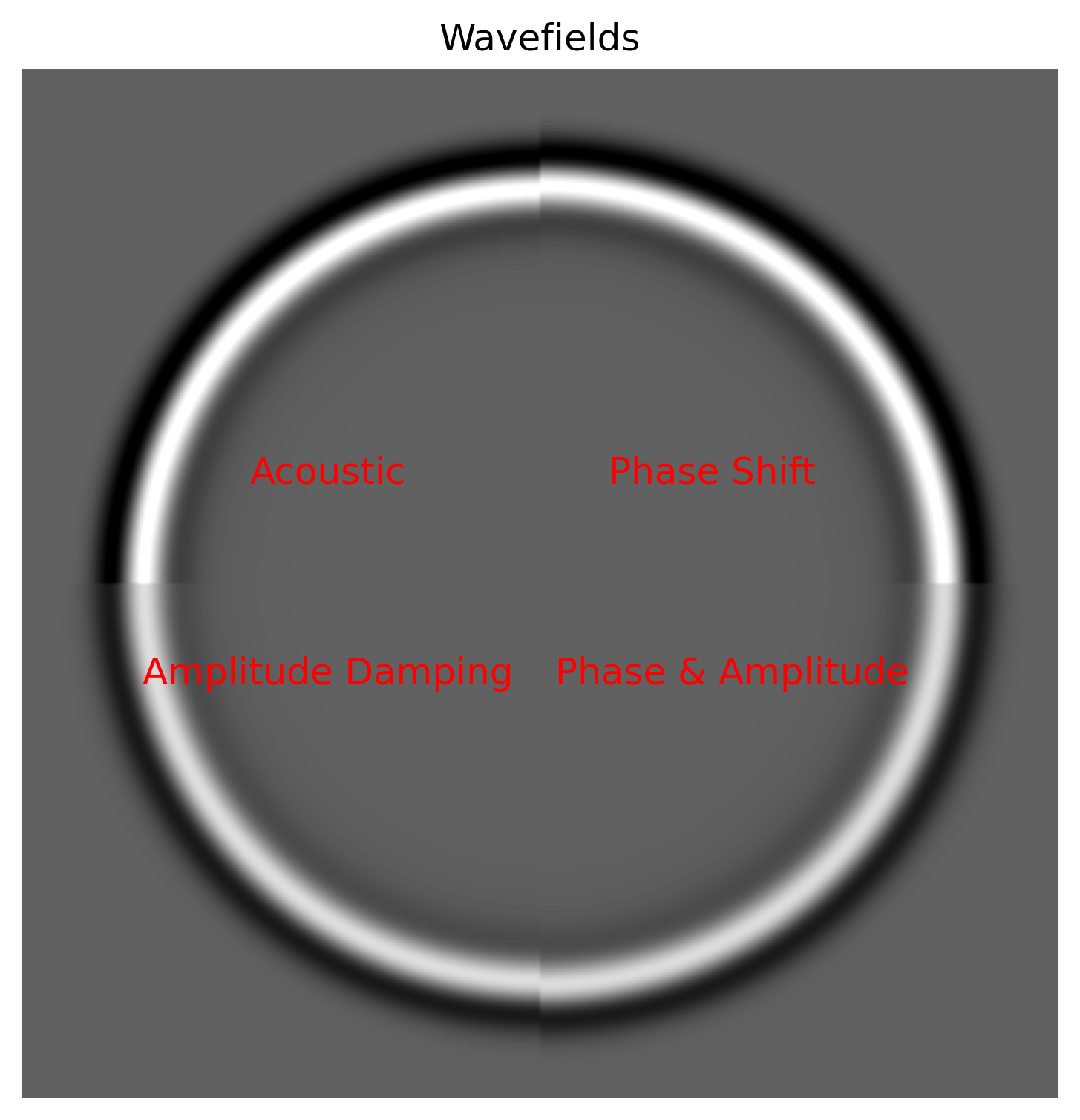}
\caption{Attenuation and dispersion in acoustic wave equations.}
\label{fig:attenuation_dispersion}
\end{figure}

Figure~\ref{fig:attenuation_dispersion} illustrates the effects of attenuation and dispersion in acoustic wave equations. The wavefield is divided into four parts: the top-left shows the original acoustic wavefield, the top-right shows the wavefield after applying a phase shift, the bottom-left shows the wavefield after applying amplitude attenuation, and the bottom-right shows the wavefield after applying both phase shift and attenuation. The differences in wave propagation characteristics are clearly observed.

\subsection{Elastic Wave Equation}

The velocity-stress elastic wave equation \citep{elastic} can be expressed as
\begin{equation}
\begin{aligned}
    \rho \frac{\partial v_x}{\partial t} &= \frac{\partial \tau_{xx}}{\partial x} + \frac{\partial \tau_{xz}}{\partial z}, \\
    \rho \frac{\partial v_z}{\partial t} &= \frac{\partial \tau_{xz}}{\partial x} + \frac{\partial \tau_{zz}}{\partial z}, \\
    \frac{\partial \tau_{xx}}{\partial t} &= (\lambda + 2\mu)\frac{\partial v_x}{\partial x} + \lambda \frac{\partial v_z}{\partial z}, \\
    \frac{\partial \tau_{zz}}{\partial t} &= (\lambda + 2\mu)\frac{\partial v_z}{\partial z} + \lambda \frac{\partial v_x}{\partial x}, \\
    \frac{\partial \tau_{xz}}{\partial t} &= \mu ( \frac{\partial v_x}{\partial z} + \frac{\partial v_z}{\partial x} ).
\end{aligned}
\end{equation}
where $\rho$ is the density, $v_x$ and $v_z$ are the particle velocities in the $x$ and $z$ directions, respectively, $\tau_{xx}$, $\tau_{zz}$, and $\tau_{xz}$ are the stress components, and $\lambda$ and $\mu$ are the Lamé parameters.
\subsection{Acoustic-Elastic Coupled Wave Equation}

The acoustic-elastic coupled wave equation \citep{aec} can be expressed as
\begin{equation}
\begin{aligned}
    \rho \frac{\partial v_x}{\partial t} &= \frac{\partial (\tau_{xx}^s - p^{2d})}{\partial x} + \frac{\partial \tau_{xz}^s}{\partial z}, \\
    \rho \frac{\partial v_z}{\partial t} &= \frac{\partial \tau_{xz}^s}{\partial x} + \frac{\partial (\tau_{zz}^s - p^{2d})}{\partial z}, \\
    \frac{\partial p^{2d}}{\partial t} &= -(\lambda + \mu)( \frac{\partial v_x}{\partial x} + \frac{\partial v_z}{\partial z} ) - h^p, \\
    \frac{\partial \tau_{xx}^s}{\partial t} &= \frac{\partial (\tau_{xx} + p^{2d})}{\partial t} = \mu( \frac{\partial v_x}{\partial x} - \frac{\partial v_z}{\partial z} ), \\
    \frac{\partial \tau_{zz}^s}{\partial t} &= \frac{\partial (\tau_{zz} + p^{2d})}{\partial t} = \mu ( \frac{\partial v_z}{\partial z} - \frac{\partial v_x}{\partial x} ), \\
    \frac{\partial \tau_{xz}^s}{\partial t} &= \mu ( \frac{\partial v_x}{\partial z} + \frac{\partial v_z}{\partial x} ).
\end{aligned}
\end{equation}

where $\rho$ is the density, $v_x$ and $v_z$ are the particle velocities in the $x$ and $z$ directions, respectively, $\tau_{xx}^s$, $\tau_{zz}^s$, and $\tau_{xz}^s$ are the stress components, $p^{2d}$ is the pressure wavefield, $\lambda$ and $\mu$ are the Lamé parameters, and $h^p$ is the source term.

\begin{figure}[h!]
\centering
\includegraphics[width=0.8\textwidth]{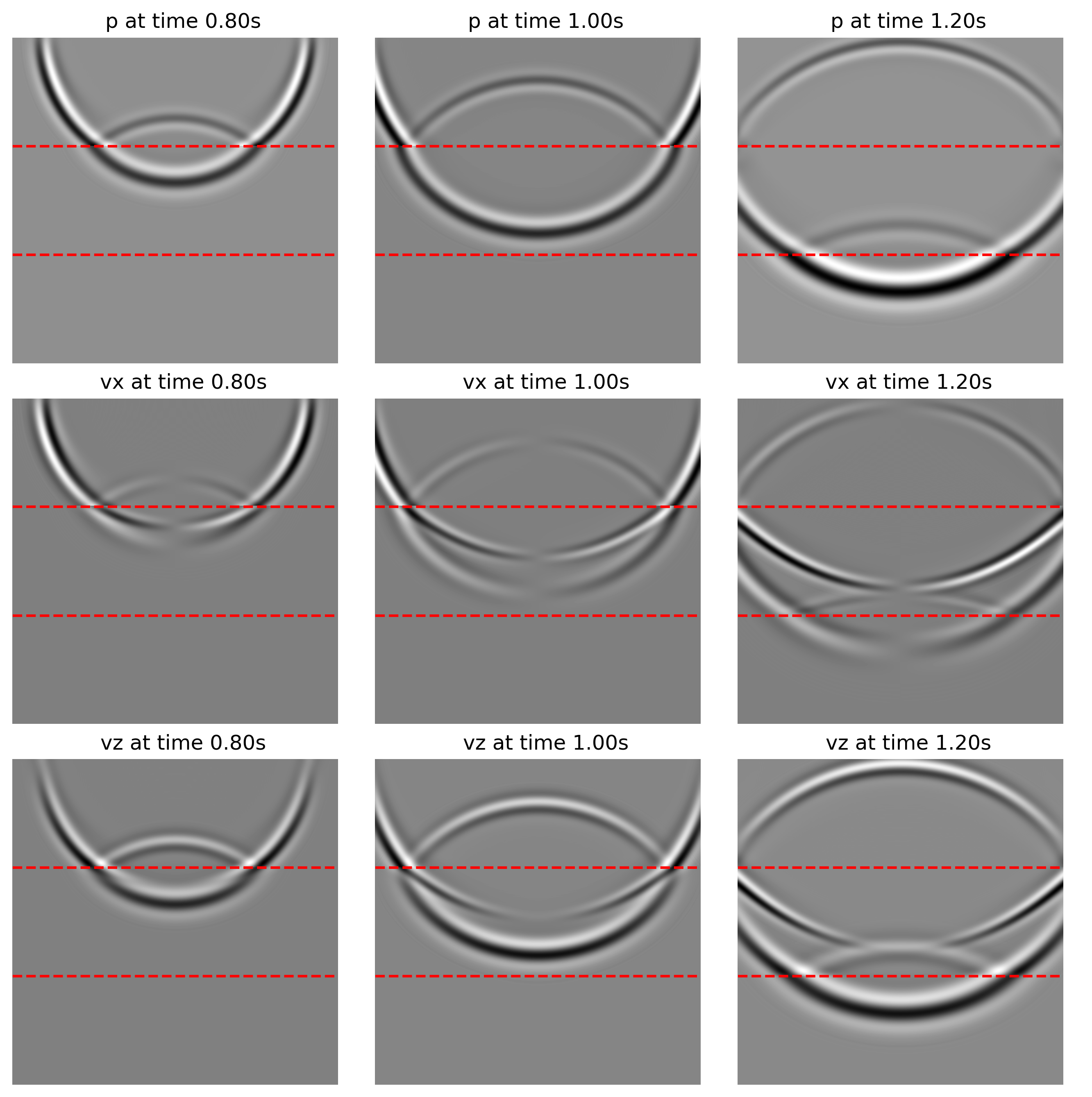}
\caption{Wavefields of $P$ (first row), $v_x$ (second row) and $v_z$ (thrid row) at different time steps.}
\label{fig:aec_wavefields}
\end{figure}

\subsubsection{Example}

The AEC is a combination of the first-order stress-velocity acoustic wave equation and the elastic wave equation. The P component in AEC can also be generated by the summation of the normal stress components $\tau_{xx}^s$ and $\tau_{zz}^s$ from the elastic wave equation. We use a three-layer model to illustrate the AEC wavefields. The P-wave velocities are 1500 $m/s$, 2000 $m/s$, and 2500 $m/s$ for the first, second, and third layers, respectively. The S-wave velocities are 0.0 $m/s$, 1200 $m/s$, and 1400 $m/s$ for the first, second, and third layers, respectively. The densities are 1000.0 $kg/m^3$, 1600 $kg/m^3$, and 2000 $kg/m^3$ for the first, second, and third layers, respectively. The source is a Ricker wavelet with a frequency of 10 Hz. The wavefields are shown in Figure~\ref{fig:aec_wavefields}.

\subsection{Pseudo Elastic Wave Equation}

The pseudo elastic wave equation \citep{pseudoelastic} can be expressed as

\begin{equation}
\begin{aligned}
\frac{\partial \tau_{xx}}{\partial t} &= (\lambda + 2 \mu) \frac{\partial}{\partial x} \left( \frac{\partial^2 v_x^{\dagger}}{\partial x^2} + \frac{\partial^2 v_z^{\dagger}}{\partial x \partial z} \right) 
+ \lambda \frac{\partial}{\partial z} \left( \frac{\partial^2 v_x^{\dagger}}{\partial x \partial z} + \frac{\partial^2 v_z^{\dagger}}{\partial z^2} \right), \\
\\
\frac{\partial \tau_{zz}}{\partial t} &= (\lambda + 2 \mu) \frac{\partial}{\partial z} \left( \frac{\partial^2 v_x^{\dagger}}{\partial x \partial z} + \frac{\partial^2 v_z^{\dagger}}{\partial z^2} \right) 
+ \lambda \frac{\partial}{\partial x} \left( \frac{\partial^2 v_x^{\dagger}}{\partial x^2} + \frac{\partial^2 v_z^{\dagger}}{\partial x \partial z} \right), \\
\\
\frac{\partial \tau_{xz}}{\partial t} &= \mu \left( \frac{\partial}{\partial x} \left( \frac{\partial^2 v_x^{\dagger}}{\partial x \partial z} + \frac{\partial^2 v_z^{\dagger}}{\partial z^2} \right) 
+ \frac{\partial}{\partial z} \left( \frac{\partial^2 v_x^{\dagger}}{\partial x^2} + \frac{\partial^2 v_z^{\dagger}}{\partial x \partial z} \right) \right), \\
\\
\rho \frac{\partial v_x}{\partial t} &= \frac{\partial \tau_{xx}}{\partial x} + \frac{\partial \tau_{xz}}{\partial z}, \\
\\
\rho \frac{\partial v_z}{\partial t} &= \frac{\partial \tau_{zz}}{\partial z} + \frac{\partial \tau_{xz}}{\partial x}, \\
\\
v_x^{\dagger} &= \frac{v_x}{\frac{\partial^2}{\partial x^2} + \frac{\partial^2}{\partial z^2}}, 
\quad
v_z^{\dagger} = \frac{v_z}{\frac{\partial^2}{\partial x^2} + \frac{\partial^2}{\partial z^2}}.
\end{aligned}
\end{equation}
where $\tau_{xx}$, $\tau_{zz}$, and $\tau_{xz}$ are the stress components, $\rho$ is the density, and $f_x$ and $f_z$ are the source terms in the $x$ and $z$ directions, respectively. The equations that contains laplace operator $\frac{\partial^2}{\partial x^2} + \frac{\partial^2}{\partial z^2}$ can be solved by pseudo spectral method, which is 
\begin{equation}
\begin{aligned}
v_x^{\dagger} &= \mathcal{F}^{-1} ( -\frac{\mathcal{F}(v_x)}{k_x^2 + k_z^2} ), \\
v_z^{\dagger} &= \mathcal{F}^{-1} ( -\frac{\mathcal{F}(v_z)}{k_x^2 + k_z^2} ), \\
\end{aligned}
\end{equation}
where $\mathcal{F}$ and $\mathcal{F}^{-1}$ are the Fourier transform and inverse Fourier transform, respectively, and $k_x$ and $k_z$ are the wavenumber components in the $x$ and $z$ directions, respectively.

\subsubsection{Example}
\begin{figure}[h!]
\centering
\includegraphics[width=0.8\textwidth]{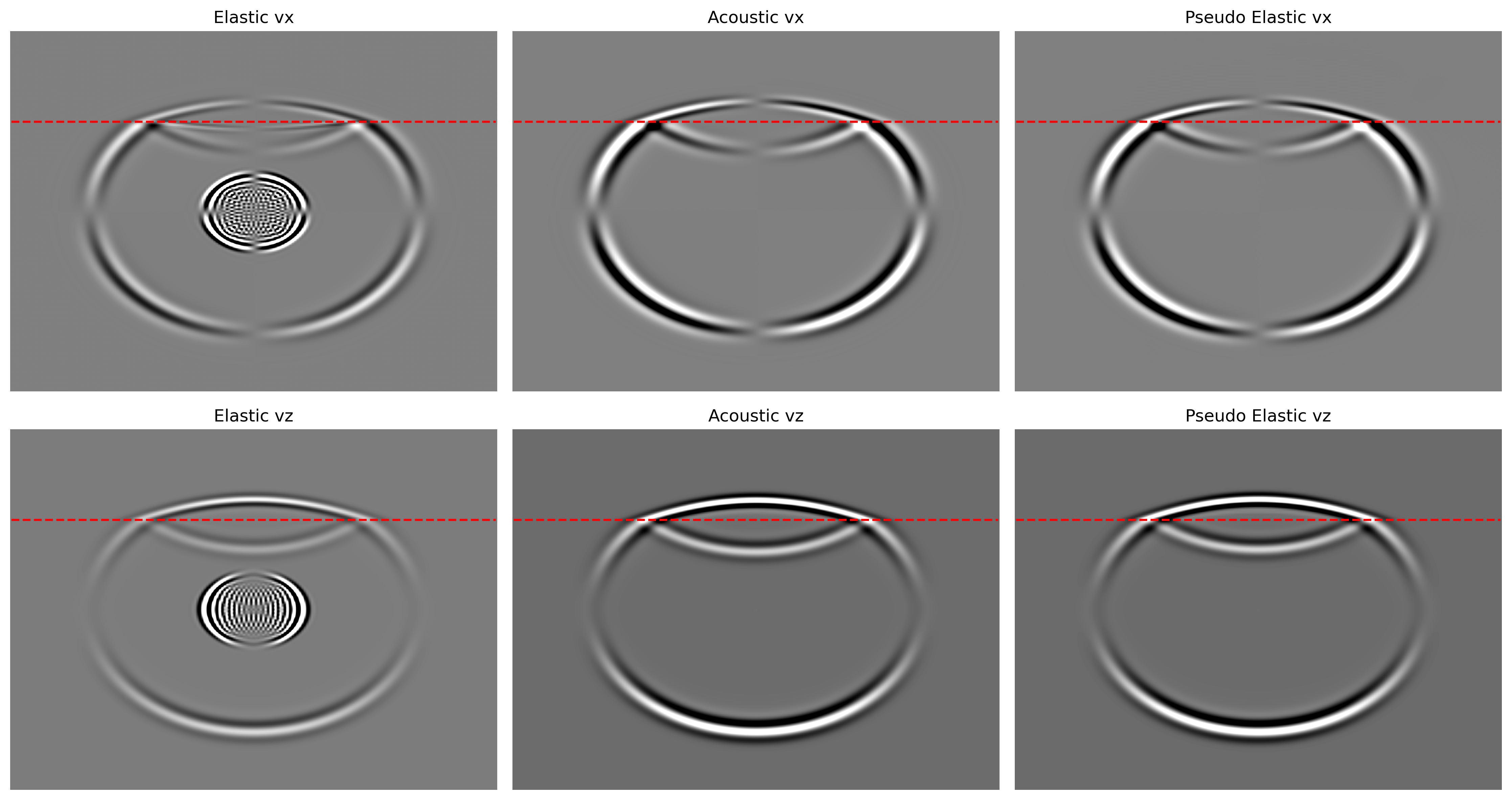}
\caption{Wavefields of the pseudo elastic wave equation.}
\label{fig:pseudo_elastic_wavefields}
\end{figure}

For comparing the wavefields, we use elastic, acoustic, and pseudo-elastic wave equations to generate the wavefields. A two-layer model with a P-wave velocity of $300$ m/s, an S-wave velocity of $0$ m/s, and a density of $1250$ kg/m$^3$ for the first layer, and a P-wave velocity of $500$ m/s, an S-wave velocity of $150$ m/s, and a density of $1800$ kg/m$^3$ for the second layer is used. The source is a Ricker wavelet with a frequency of $8$ Hz. The wavefields are shown in Figure~\ref{fig:pseudo_elastic_wavefields}. We can see that the pseudo-elastic wave equation generates wavefields similar to those of the acoustic wave equation, whereas the elastic wave equation exhibits dispersion due to the low S-wave velocity in the second layer.

\subsection{Acoustic isotropic LSRTM}

The background wavefield $P$ and perturbed wavefield $\delta P$ can be solved by the following equations \citep{acousticlsrtm}:
\begin{equation}
\begin{aligned}
&\frac{\partial^2 P}{\partial t^2} = v_{p0}^2 \left( \frac{\partial^2 P}{\partial x^2} + \frac{\partial^2 P}{\partial z^2} \right),\\
&\frac{\partial^2 \delta P}{\partial t^2} = v_{p0}^2 \left( \frac{\partial^2 \delta P}{\partial x^2} + \frac{\partial^2 \delta P}{\partial z^2} \right) + m \left( \frac{\partial^2 P}{\partial x^2} + \frac{\partial^2 P}{\partial z^2} \right),
\end{aligned}
\end{equation}

where $v_{p0}$ is the background P-wave velocity, and $m$ is the reflectivity, which is defined as $m=2\delta v_p/v_{p0}$.
\subsection{Elastic isotropic LSRTM}
The elastic LSRTM contains two parts: the background wavefield and the perturbed wavefields. The forward wavefields can be modelled by the elastic wave equation \citep{elastic}, which can be expressed as

\begin{equation}
\begin{aligned}
    \rho \frac{\partial v_x}{\partial t} &= \frac{\partial \tau_{xx}}{\partial x} + \frac{\partial \tau_{xz}}{\partial z}, \\
    \rho \frac{\partial v_z}{\partial t} &= \frac{\partial \tau_{xz}}{\partial x} + \frac{\partial \tau_{zz}}{\partial z}, \\
    \frac{\partial \tau_{xx}}{\partial t} &= (\lambda + 2\mu)\frac{\partial v_x}{\partial x} + \lambda \frac{\partial v_z}{\partial z}, \\
    \frac{\partial \tau_{zz}}{\partial t} &= (\lambda + 2\mu)\frac{\partial v_z}{\partial z} + \lambda \frac{\partial v_x}{\partial x}, \\
    \frac{\partial \tau_{xz}}{\partial t} &= \mu ( \frac{\partial v_x}{\partial z} + \frac{\partial v_z}{\partial x} ).
\end{aligned}
\end{equation}

The perturbed wavefields \citep{elasticlsrtm} can be written as
\begin{equation}
\begin{aligned}
    \rho \frac{\partial \delta v_x}{\partial t} &= \frac{\partial \delta \tau_{xx}}{\partial x} + \frac{\partial \delta \tau_{xz}}{\partial z}, \\
    \rho \frac{\partial \delta v_z}{\partial t} &= \frac{\partial \delta \tau_{xz}}{\partial x} + \frac{\partial \delta \tau_{zz}}{\partial z}, \\
    \frac{\partial \delta \tau_{xx}}{\partial t} &= (\lambda + 2\mu)\frac{\partial \delta v_x}{\partial x} + \lambda \frac{\partial \delta v_z}{\partial z} + (\delta\lambda + 2\delta\mu)\frac{\partial v_x}{\partial x} + \delta\lambda \frac{\partial v_z}{\partial z}, \\
    \frac{\partial \delta  \tau_{zz}}{\partial t} &= (\lambda + 2\mu)\frac{\partial\delta v_z}{\partial z} + \lambda \frac{\partial\delta  v_x}{\partial x} + (\delta\lambda + 2\delta\mu)\frac{\partial v_z}{\partial z} + \delta\lambda \frac{\partial  v_x}{\partial x}, \\
    \frac{\partial \tau_{xz}}{\partial t} &= \mu ( \frac{\partial v_x}{\partial z} + \frac{\partial v_z}{\partial x} ) + \delta\mu ( \frac{\partial v_x}{\partial z} + \frac{\partial v_z}{\partial x} ).
\end{aligned}
\end{equation}
where $\delta v_x$ and $\delta v_z$ are the perturbed particle velocities, $\delta \tau_{xx}$, $\delta \tau_{zz}$, and $\delta \tau_{xz}$ are the perturbed stress components, and $\delta\lambda$ and $\delta\mu$ are the perturbations of the Lamé parameters.
\subsection{Acoustic-Elastic Coupled LSRTM}
The background wavefields can be modelled by the acoustic-elastic coupled wave equation \citep{aec}, which can be expressed as
\begin{equation}
\begin{aligned}
    \rho \frac{\partial v_x}{\partial t} &= \frac{\partial (\tau_{xx}^s - p^{2d})}{\partial x} + \frac{\partial \tau_{xz}^s}{\partial z}, \\
    \rho \frac{\partial v_z}{\partial t} &= \frac{\partial \tau_{xz}^s}{\partial x} + \frac{\partial (\tau_{zz}^s - p^{2d})}{\partial z}, \\
    \frac{\partial p^{2d}}{\partial t} &= -(\lambda + \mu)( \frac{\partial v_x}{\partial x} + \frac{\partial v_z}{\partial z} ) - h^p, \\
    \frac{\partial \tau_{xx}^s}{\partial t} &= \frac{\partial (\tau_{xx} + p^{2d})}{\partial t} = \mu( \frac{\partial v_x}{\partial x} - \frac{\partial v_z}{\partial z} ), \\
    \frac{\partial \tau_{zz}^s}{\partial t} &= \frac{\partial (\tau_{zz} + p^{2d})}{\partial t} = \mu ( \frac{\partial v_z}{\partial z} - \frac{\partial v_x}{\partial x} ), \\
    \frac{\partial \tau_{xz}^s}{\partial t} &= \mu ( \frac{\partial v_x}{\partial z} + \frac{\partial v_z}{\partial x} ).
\end{aligned}
\end{equation}
and the perturbed wavefields can be written as \citep{aeclsrtm}
\begin{equation}
\begin{aligned}
    \rho \frac{\partial\delta v_x}{\partial t} &= \frac{\partial (\delta\tau_{xx}^s - \delta p^{2d})}{\partial x} + \frac{\partial\delta \tau_{xz}^s}{\partial z}, \\
    \rho \frac{\partial\delta v_z}{\partial t} &= \frac{\partial\delta \tau_{xz}^s}{\partial x} + \frac{\partial (\delta\tau_{zz}^s - \delta p^{2d})}{\partial z}, \\
    \frac{\partial\delta p^{2d}}{\partial t} &= -(\lambda + \mu)( \frac{\partial\delta v_x}{\partial x} + \frac{\partial\delta v_z}{\partial z} ) -(\delta\lambda + \delta\mu)( \frac{\partial v_x}{\partial x} + \frac{\partial v_z}{\partial z} ), \\
    \frac{\partial\delta \tau_{xx}^s}{\partial t} &= \mu( \frac{\partial\delta v_x}{\partial x} - \frac{\partial\delta v_z}{\partial z} )+\delta\mu( \frac{\partial v_x}{\partial x} - \frac{\partial v_z}{\partial z} ), \\
    \frac{\partial\delta \tau_{zz}^s}{\partial t} &= \mu ( \frac{\partial\delta v_z}{\partial z} - \frac{\partial\delta v_x}{\partial x} )+\delta\mu( \frac{\partial v_z}{\partial z} - \frac{\partial v_x}{\partial x} ), \\
    \frac{\partial\delta \tau_{xz}^s}{\partial t} &= \mu ( \frac{\partial\delta v_x}{\partial z} + \frac{\partial\delta v_z}{\partial x} )+\delta\mu ( \frac{\partial v_x}{\partial z} + \frac{\partial v_z}{\partial x} ).
\end{aligned}
\end{equation}

where $\delta v_x$ and $\delta v_z$ are the perturbed particle velocities, $\delta \tau_{xx}^s$, $\delta \tau_{zz}^s$, and $\delta \tau_{xz}^s$ are the perturbed stress components, $\delta p^{2d}$ is the perturbed pressure wavefield, and $\delta \lambda$ and $\delta \mu$ are the perturbations of the Lamé parameters. (In \cite{aeclsrtm}, these equations were simplified by replacing $\tau_{xx}$ with $-\tau_{zz}$; however, in this report, we retain the original form.)

\section{Conclusion}
In conclusion, Sweep provides a powerful, flexible, and extensible platform for seismic wave modeling and inversion. By supporting a wide range of wave equation solvers and offering native automatic differentiation, it streamlines the implementation of advanced inversion techniques such as FWI and LSRTM. Its modular design, support for multi-GPU computing, and seamless integration with neural networks make it well-suited for tackling complex and large-scale seismic inverse problems. Sweep thus serves as a unified solution for both traditional and emerging approaches in computational geophysics.

\section{Acknowledgements}
The authors thank KAUST and the Deepwave consortium for their support. For computer time, this research used the resources of the Supercomputing Laboratory at KAUST in Thuwal, Saudi Arabia.

\bibliographystyle{unsrt}
\bibliography{sample}

\end{document}